\PassOptionsToPackage{final}{graphicx}
\documentclass[sigconf,natbib=false]{acmart}

\usepackage[backend=biber,bibencoding=utf8,style=ACM-Reference-Format,natbib,mincitenames=1,maxcitenames=2,autocite=inline,sortcites,labelnumber]{biblatex}
\DeclareFieldFormat*{citetitle}{\emph{#1}}

\hypersetup{draft=false}

\usepackage[obeyDraft,colorinlistoftodos,prependcaption]{todonotes}

\makeatletter

\usepackage{comment}
\let\wfs@comment@comment\comment%
\let\comment\@undefined%

\usepackage{changes}
\let\wfs@changes@comment\comment%
\let\comment\@undefined%

\newcommand\comment{%
    \ifthenelse{\equal{\@currenvir}{comment}}
    {\wfs@comment@comment}
    {\wfs@changes@comment}%
}

\makeatother

\definechangesauthor[name={Raffi Khatchadourian}]{RK}

\definechangesauthor[name={Yiming Tang},color=purple]{YT}

\usepackage[group-separator={,},group-minimum-digits={3}]{siunitx}
\DeclareSIUnit[number-unit-product = ]\percent{\char`\%}

\usepackage{nicefrac}
\usepackage{textcomp}
\usepackage{soul}
\usepackage{mathrsfs}
\usepackage{amsmath}
\usepackage{amsfonts}
\usepackage{mathtools}
\usepackage{numname}
\usepackage[capitalize]{cleveref}

\crefname{listing}{listing}{listings}
\Crefname{listing}{Listing}{Listings}
\crefname{sublisting}{listing}{listings}
\Crefname{sublisting}{Listing}{Listings}

\usepackage{ellipsis}

\usepackage{algorithm}
\usepackage{algorithmic}

\crefname{ALC@unique}{line}{lines}
\Crefname{ALC@unique}{Line}{Lines}

\crefname{paragraph}{section}{sections}
\Crefname{paragraph}{Section}{Sections}

\graphicspath{{./graphics/}} 

\usepackage[shortlabels,inline]{enumitem}

\usepackage[draft=false, frozencache]{minted}
\newminted{diff}{
	style=trac,
}
\newminted{java}{}
\newmintinline{java}{}
\setmintedinline{
    style=bw,
    fontsize=\small}
\setminted{%
    fontsize=\footnotesize,
    autogobble,
    breaklines,
    resetmargins,
    breakindentnchars=4,
    tabsize=4,
    breakafter=.,
    breaksymbol=,
    breakautoindent,
    breakaftersymbolpre=,
    escapeinside=||}

\newcommand{\ji}[1]{\javainline{#1}}

\usepackage{booktabs}
\usepackage{tabulary}
\usepackage{threeparttable}
\usepackage{multirow}

\theoremstyle{definition}

\newcommand{\JUL}{\href{https://docs.oracle.com/en/java/javase/17/docs/api/java.logging/java/util/logging/package-summary.html}{\ji{Java Util Logging}}}
\newcommand{\slf}{\href{http://www.slf4j.org/}{Slf4j}}

\newcommand{\jenkins}{\href{https://github.com/jenkinsci/jenkins}{Jenkins}}
\newcommand{\selenium}{\href{https://github.com/SeleniumHQ/selenium}{selenium}}

\setcopyright{acmcopyright}
\copyrightyear{2022}
\acmYear{2022}
\setcopyright{acmcopyright}\acmConference[ICSE '22 Companion]{44th International Conference on Software Engineering Companion}{May 21--29, 2022}{Pittsburgh, PA, USA}
\acmBooktitle{44th International Conference on Software Engineering Companion (ICSE '22 Companion), May 21--29, 2022, Pittsburgh, PA, USA}
\acmPrice{15.00}
\acmDOI{10.1145/3510454.3516838}
\acmISBN{978-1-4503-9223-5/22/05}

\begin{document}

\title{A Tool for Rejuvenating Feature Logging Levels via Git Histories and Degree of Interest}

\author{Yiming Tang}
\orcid{0000-0003-2378-8972} 
\affiliation{
    \institution{Concordia University}
    \city{Montreal}
    \country{Canada}
}
\email{t_yiming@encs.concordia.ca}

\author{Allan Spektor}
\affiliation{
    \institution{CUNY Hunter College}
    \city{New York}
    \state{NY}
    \country{USA}
}
\email{allan.spektor03@myhunter.cuny.edu}

\author{Raffi Khatchadourian}
\orcid{0000-0002-7930-0182} 
\affiliation{
    \institution{CUNY Hunter College}
    \city{New York}
    \state{NY}
    \country{USA}
}
\email{raffi.khatchadourian@hunter.cuny.edu}

\author{Mehdi Bagherzadeh}
\affiliation{
    \institution{Oakland University}
    \city{Rochester}
    \state{MI}
    \country{USA}
}
\email{mbagherzadeh@oakland.edu}



\begin{abstract}

    Logging is a significant programming practice. Due to the highly transactional nature of modern software applications, massive amount of logs are generated every day, which may overwhelm developers. Logging information overload can be dangerous to software applications. Using log levels, developers can print the useful information while hiding the verbose logs during software runtime. As software evolves, the log levels of logging statements associated with the surrounding software feature implementation may also need to be altered. Maintaining log levels necessitates a significant amount of manual effort. In this paper, we demonstrate an automated approach that can rejuvenate feature log levels by matching the interest level of developers in the surrounding features. The approach is implemented as an open-source Eclipse plugin, using two external plug-ins (JGit and Mylyn). It was tested on 18 open-source Java projects consisting of $\sim$3 million lines of code and $\sim$4K log statements. Our tool successfully analyzes 99.22\% of logging statements, increases log level distributions by $\sim$20\%, and increases the focus of logs in bug fix contexts $\sim$83\% of the time. For further details, interested readers can watch our demonstration video (\url{https://www.youtube.com/watch?v=qIULoAXoDv4}).

\end{abstract}


	

\maketitle


\section{Introduction}\label{sec:intro}

Logging is a widely used programming practice for recording software system information during runtime~\cite{Yuan2012,Li2017a,Li2017}. The significance of logging is incontrovertible, as the runtime information stored in logs is used by developers for a variety of purposes, such as monitor processes~\cite{Rozinat2005}, transferring knowledge~\cite{Kabinna2018}, and error detection~\cite{Zeng2019,Syer2013}. 

With the evolution of modern software development, modern software now has the potential to analyze massive amounts of data on a daily basis. Due to such highly transactional nature of modern software, it can generate massive amounts of logs every day, which may overwhelm developers. Many prior studies have highlighted the dangers posed by logging information overload. For example, \citet{Yuan2012} indicate that excessive logging may deliver too much noise which inhibits error detection. According to \citet{Fu2014}, information overload might result in additional hardware, development, and maintenance expenses, as well as redundant log data.

Many mainstream programming languages come with logging frameworks (e.g., \JUL\ in Java and \href{https://docs.python.org/3/library/logging.html}{\ji{logging}} in Python) that help developers standardize logging practices. Using logging frameworks, developers can place logging statements into source code for generating runtime logs. Typical logging statements include log levels, which allow developers to specify which logs are visible during software run-time, while hiding the verbose logs. Specifically, logging frameworks treat a certain log level as a default verbosity level, enabling logging statements with log levels greater than or equal to it to emit logs at runtime. For instance, the logging statement below is taken from the \href{https://docs.oracle.com/en/java/javase/17/docs/api/java.logging/java/util/logging/Logger.html}{JUL documentation} and its log level is \ji{FINER}. If the default verbosity level is set to \ji{FINER}, this logging statement could print log messages at runtime. Therefore, setting appropriate log levels can aid developers in receiving useful log information for further development and testing.

\begin{javacode} 
logger.log(Level.FINER, DiagnosisMessages::systemHealthStatus);
\end{javacode}

As software evolves, new software features are introduced, while some existing software features are enhanced or suppressed. The log levels of logging statements associated with the surrounding software feature implementation, referred to as \textit{feature log levels}, may also need to be altered, as these logging statements are unable to offer developers with the most up-to-date information that is no longer of interest to them. 
For example, if a software feature is suppressed, its relevant logs are no longer appealing to developers and should be concealed during software runtime to assist developers in receiving information more effectively.
In ideal situations, the levels of feature logs should be raised when more developers become interested in surrounding software features, and vice versa, resulting in more valuable log information being displayed and less useful information being suppressed during software runtime. To identify feature logs, we established a set of innovative heuristics based on first-hand developer interactions, which can be found on the tool's main menu and are explored in more detail in \cref{sec:heur_design}.

A prior study~\cite{Yuan2012} discloses that developers often fail to set log levels appropriately the first time, and then alter them afterward. Evolving log levels necessitates a significant amount of manual effort. To the best of our knowledge, there are no automated existing approaches for maintaining log levels by taking into account the evolution of surrounding software features. Therefore, we propose an automated approach, namely \textbf{REFELL}, that can \underline{re}juvenate \underline{fe}ature \underline{l}og \underline{l}evels by matching the interest level of developers in the surrounding features. 
 Our approach attempts to aid developers in effectively getting log information as software evolves, because appropriate log levels can prevent developers from receiving too much or too little software run-time information.
The approach analyzes Git histories for code changes, then adapts Mylyn's \underline{D}egree \underline{o}f \underline{I}nterest (DOI) model~\cite{Kersten2005} to gauge developers' interest in the software source code surrounding the logging statements based on the retrieved code changes. Mylyn~\cite{EclipseFoundation2020} is an Eclipse plugin whose basic algorithm is the DOI model, which enables program elements with more frequent and recent interactions to be highlighted more prominently, and vice versa. The approach correlates such interests with feature log levels and suggests new log levels if mismatches are discovered.

The approach is implemented as an open-source Eclipse plugin, which developers can download and install using an Eclipse update site link.\footnote{\url{https://git.io/J17UC}} The project source code is also publicly available on GitHub.\footnote{\url{https://git.io/JMTNW}} To explore the approach's capabilities in real-life applications, we conducted experiments on 18 open-source Java projects. Our tool examines $\sim$4K logging statements, improves log level distributions by $\sim$20\%, and increases the focus of logs in bug fix contexts by 83\%. Several pull requests were also incorporated into prominent and well-known open-source projects.

The corresponding full technical paper appeared in \textit{Science of Computer Programming}~\cite{Tang2021}. The full paper contains further information, such as the approach introduction, evaluation design, and discussion.

\section{Envisioned users}

The users we expect to attract are the developers and even testers of large software systems. These software systems are constantly updated and can generate a large number of logs per day. Analyzing these logs to obtain useful information necessitates a significant amount of human effort. Our tool can assist them in receiving run-time log information more efficiently.

\section{How the tool is used}

We provide a user-friendly, easy-to-use tool to developers. Users should first install our tool in Eclipse and create a new Mylyn task before using it. The Mylyn task needs to be activated. After that, users only need to choose the assessed projects and click on the \textsc{rejuvenate a log level} command via Quick Assess.

\begin{figure}[h]
    \includegraphics[width=\linewidth]{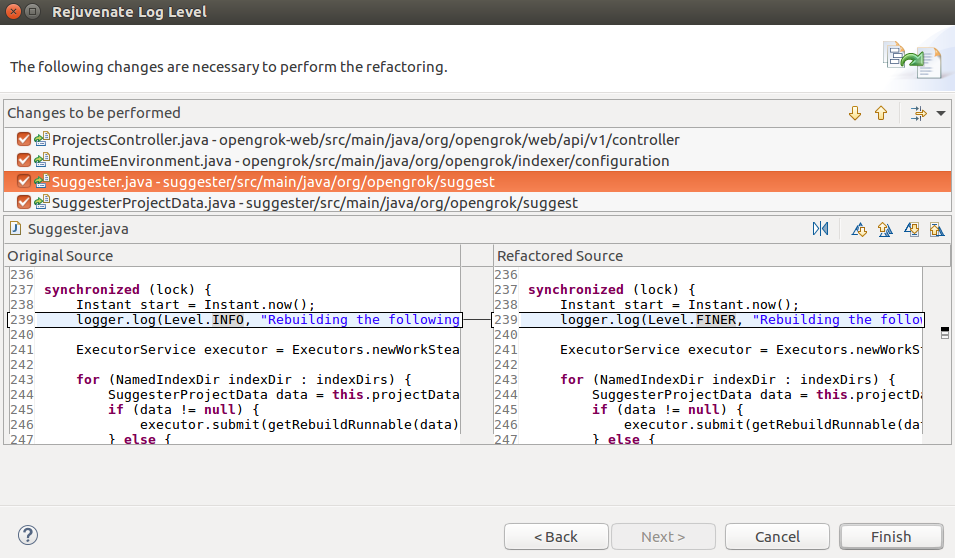}
    \caption{Screenshot of the preview wizard for \textsc{REFELL}.}\label{fig:preview}
\end{figure}

The main menu of the tool includes the heuristics listed in Table~\ref{tab:heuristics}. After choosing heuristics, the tool analyzes source code, and a preview dialog box appears, as shown in Figure~\ref{fig:preview}. In this dialog box, users can choose the source file to view and check all transformations in this file. If users agree with the log level transformations, they can perform log level transformations by clicking on the ``Finish'' button. For more information, interested readers could watch our demonstration video.

\section{Software Engineering Challenges}

As we stated in \cref{sec:intro}, our tool expects to reconcile the software log information overload. With the help of our tool, the log information that users are more interested in has a better chance of being printed during software run-time, and vice versa. The following is a log level transformation that our tool recommended and was approved by \jenkins\ developers.
\begin{diffcode}
    -	LOGGER.log(INFO,"{0} main build action completed:{1}"..);
    +	LOGGER.log(FINEST,"{0} main build action completed:{1}..");
\end{diffcode}
In this case, the original log level is \ji{info}, indicating that when developers chose this log level, they were interested in the log information provided by this logging statement. However, after that, the project went through a long development period, and the log information from this logging statement was no longer as interesting to developers as it once was. Our tool integrates with the Eclipse Mylyn plug-in and can track developers' interest in software features surrounding logging statements. The tool discovered that developers were less interested in the software features surrounding this logging statement. In addition, this logging statement was associated with its surrounding software features. Therefore, our tool suggested lowering the log level in this case, which was agreed upon by developers. \jenkins\ developers stated “[it is p]robably a good idea: [i]t’s time we started removing this from the general system log~\cite{pr-for-jenkins}.”

\section{Approach and Implementation}

\subsection{Architecture and Dependencies}

\begin{figure}[h]
    \includegraphics[width=0.75\linewidth]{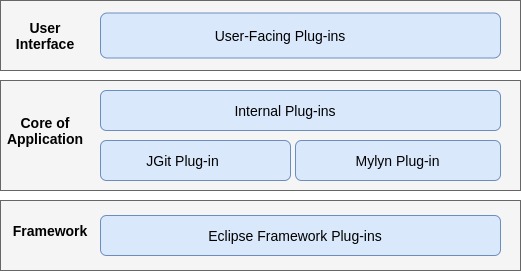}
    \caption{Software Architecture for \textsc{REFELL}.}\label{fig:arch}
\end{figure}

Our tool's design is depicted in \cref{fig:arch}, which consists mostly of three layers. The top layer is the user interface layer that accepts source code from software projects as input. The medium layer implements the approach's primary functionality.  It leverages two external plug-ins: JGit for historical source code change extraction and Mylyn for measuring developer interests. Our tool mines every project's Git history to extract a collection of source code modifications. For each source code modification, the tool creates an interaction event as a Mylyn input. Mylyn could then automatically measure developers' interest in program elements (methods) enclosing logging statements. The developer's interest in a software feature is quantified as a real dubbed the DOI value. The DOI ranges of features are later split by subtracting the smallest DOI value from the largest, and then dividing the result by the number of available logging levels. 
As a result, each DOI partition is associated with a log level. 
Therefore, the tool can predict a log level for each log level based on developers' interests in the surrounding software feature, and recommend a new log level if the existing log level and the anticipated log level differ. 
Currently, every log level corresponds to the same size DOI partition. We chose same size since it is the most straightforward and intuitive strategy, and we will apply Machine Learning technologies in the future to enhance our algorithms.
The bottom layer in \cref{fig:arch} is the basic foundation of this tool, which can provide Eclipse plug-in development support for a transformation tool.

\subsection{Integration with Mylyn}

In this section, more information on how historical source code modifications are transformed to interaction events as Mylyn input is provided. For each software feature change (method change), the tool sequentially builds a Mylyn interaction event with kind \ji{EDIT}, indicating that this interaction event is for editing source code. Mylyn might construct many types of interaction events, such as \ji{SELECTION}, for a common Mylyn task context. However, because Git commits cannot hold as much information as all finer-grained interaction event types defined in Mylyn, the tool has to ignore the others and only consider the \ji{EDIT} type. In the future, we would like to merge the existing Mylyn task context with the simulative task context created from Git history. Existing task contexts contain a wide range of interaction events that could facilitate log transformations.

Developers' interest in a program element is reflected by the DOI value, with a higher DOI value indicating greater developer interest. DOI values are higher for program elements with more frequent and recent interactions than for program elements with fewer frequent and recent interactions. A developers' interaction could raise the DOI value of the relevant program element while lowering DOI values for other program elements that are unaffected by the interaction. In the DOI model, decreasing DOI values is referred to as \textit{decay}. Due to the presence of decay, when our tool integrates with Mylyn, the developers' interest in the program elements for the very early commits is less significant and less considered than the developers' interest extracted from the most recent commits. If the analyzed project has a long Git history, the early commits have little impact on the final log level transformation. In addition, our tool provides users with the option to limit the number of analyzed Git commits to avoid the tool analyzing very ancient commits.
Further details about the examined Git histories can be found in our full technical paper~\cite{Tang2021}.

In Mylyn, negative DOI values indicate uninterestingness. In Eclipse, the view ``Focus on Active Task'' only displays elements with positive DOI values, while those with negative DOI values are hidden.  Our tool has the potential to generate negative DOI values. If a program element in a Mylyn active context is not visited indirectly or directly for a long period of time, its DOI value could decay many times and may become negative. As a result, we treat negative DOI values as 0. The DOI has a minimum value of 0.


Multiple projects are treated as independent projects and processed individually. The Mylyn task context is cleared after each processing. Each project's Git history needs to be traversed, so the Git history of a repository with different subprojects may be analyzed more than once. Memoization is used to eliminate duplicate analysis. Each time the tool analyzes a project, it saves some intermediate data, such as source code changes. When the tool analyzes projects that share a repository, the stored data is read directly. Therefore, a repository is only processed once for every run of the tool.

\subsection{Heuristic Design}\label{sec:heur_design}

Heuristics are designed to distinguish between feature logs and non-feature logs. We performed a pilot study by analyzing several well-known open-source projects using our initial version of the tool collecting comments from their developers in order to refine and improve our heuristics design. Table~\ref{tab:heuristics} lists all of the available heuristics in the approach. These heuristics are also displayed as transformation setting options in the tool's main menu, allowing developers to accept or reject them while using the tool. We would suggest users take all of the heuristics stated in the table~\ref{tab:heuristics} into account since these heuristics can filter out non-feature logs to avoid undesirable log level transformations.

\begin{table}[h]
    \footnotesize
    \caption{A list of heuristics to identify feature logs}\label{tab:heuristics}
    \centering
    \begin{tabular}{ l m{0.75\columnwidth} }
    \toprule
    \textbf{Heuristic} &\textbf{Details} \\
    \midrule

    \textbf{WS} & Treat particular levels as log categories. \\
    \midrule

    \textbf{LOW}& Never lower the level of logging statements:
    \begin{tabular}%
        {l}
        (a) \textbf{CTCH}: appearing within catch blocks. \\
        (b) \textbf{IFS}: immediately following branches (e.g., if, switch). \\
        (c) \textbf{KEYL}: having particular keywords in their log messages. \\
        \end{tabular}%
    \\ 
    \midrule
    \textbf{CNDS} & Never change the level of logging statements immediately following branches whose condition contains a log level. \\
    \midrule
    \textbf{KEYR} & Never raise the level of logging statements without particular keywords in their log messages. This is only applicable to critical log levels (e.g. \ji{WARNING} and \ji{SEVERE} in \JUL).\\
    \midrule
    \textbf{INH} & Only consistently transform the level of logging statements appearing in overriding methods.\\
    \midrule
    \textbf{TDIST} & Only transform the level of logging statements up to a transformation distance threshold.\\
    \bottomrule
    \end{tabular}
    \end{table}


\subsection{Log Level Extraction and Rewrite}

Currently, our tool supports two popular Java logging frameworks: \JUL\ (a Java built-in logging framework) and \slf\ (a third-party logging framework). According to API documentation~\cite{Oracle2021a,QOS.ch2021b}, log levels can occur in logging statements in two ways:
\begin{enumerate*}[(i)]
    \item log level can be passed as a method parameter of a logging statement, as seen in the code example in the \cref{sec:intro}, and
    \item the method name of a convenience method could match to a log level.
\end{enumerate*}
For example, the logging statement below is from the Slf4j documentation~\cite{QOS.ch2021b}, and its log level is \ji{info}, which is the same as its method name.

\begin{javacode}
logger.info("Temperature has risen above 50 degrees.");
\end{javacode}

The tool uses AST and its source symbol bindings to extract logging statements from source code, then uses log level string match to retrieve log levels from those statements. Logging transformations are implemented by rewriting AST\@.

\section{Evaluation}

The tool was tested on 18 open-source Java projects of various sizes and domains.
Our tool successfully assessed 99.22\% of the 3,973 logging statements, with 2,819 being non-feature logs and 1,154 being feature logs. 
The failed cases are those logging statements with log levels stored in variables that our string matching strategy cannot identify. Data-flow analysis can be used to solve those cases, but since they account for a very small number of analyzed logging statements, we leave them for future research.
Among the detected feature logs, 753 are suggested to be transformed. The tool improves log level distributions by $\sim$20\%. Among the transformed log levels, 89.51\% of the levels are lowered, implying that our tool helps developers filter out much information they are uninterested in and allowing them to concentrate on fewer features of interest.

We then conduct a bug study to see how our tool can help focus on buggy code.
We compare the log level transformations made by our tool to the ideal log level transformations. Ideally, if possible, feature log levels in buggy context (context with bug fixes) should be raised, whereas feature log levels in non-buggy context should be decreased, so that logs from logging statements in buggy context become more prominent, while logs from logging statements in non-buggy context become less prominent, allowing developers to focus on the buggy information to facilitate debugging.
According to the results, our tool raises the focus of logs in bug fix contexts $\sim$83\% of the time. The results indicate that our tool has the capacity to bring erroneous feature implementations to light and expose bugs. 

During the evaluation, we discovered that our tool still has certain limitations. For example, we may overlook some of the ``wrapped'' logging statements excluded by the \textbf{CNDS} heuristic. A ``wrapped'' logging statement is guarded by  a run-time log level check. In fact, such logging statements make up only a small portion of analyzed logging statements (only $\sim$6.3\%). For now, the tool's limitations appear to be manageable, and we will consider incorporating advanced technologies and algorithms to improve it in the future.

In addition, we perform a pull request study to assess the usefulness of our heuristic rules. As a result, pull requests have been integrated into two large and well-known open-source projects (i.e., \jenkins\ and \selenium). 
\section{Related Work}


In recent years, many research studies have been conducted on log challenges. 
\citet{Yuan2012b} implement a tool that adds appropriate logging statements into source code to enhance failure diagnosis, but this cannot solve the problem of information overload. \citet{Zhu2015} provide a logging suggestion tool which assists developers in determining where to log. However, this tool does not allow developers to alter log levels.
\citet{Chen2017} implement a tool to detect mismatches between log content and the associated log level.
However, they do not take into account how developer interest in the surrounding software features changes over time during software evolution.

\section{Conclusion \& Future Work}

In conclusion, we proposed a tool \textsc{REFELL} that can automatically rejuvenate logging statement levels by using Git histories and DOI model. 
The tool recommends a new log level if a log level does not align with developers' interests in surrounding software features.
The tool was implemented as an Eclipse plug-in by using two external Eclipse plug-ins JGit and Mylyn. It was tested on 18 open-source Java projects consisting of $\sim$3 million lines of code and $\sim$4K log statements. Our tool successfully analyzes 99.22\% of logging statements, increases log level distributions by $\sim$20\%, and increases the focus of logs in bug fix contexts $\sim$83\%of the time.

In the future, we will explore algorithm enhancement by incorporating advanced techniques such as Machine Learning based techniques and data-flow analysis. Moreover, we will explore integrating Mylyn task context.
Our work, such as automatic task context building from Git histories, may contribute back to Mylyn project, which can resolve the pending bugs in their forums~\cite{create-context}.

\printbibliography%

\end{document}